\documentclass[twocolumn,preprintnumbers,amsmath,amssymb]{revtex4}

\begin{document}
\newcommand{\gsim}{ \mathop{}_{\textstyle \sim}^{\textstyle >} }
\newcommand{\lsim}{ \mathop{}_{\textstyle \sim}^{\textstyle <} }
\newcommand{\vev}[1]{ \left\langle {#1} \right\rangle }
\newcommand{\bra}[1]{ \langle {#1} | }
\newcommand{\ket}[1]{ | {#1} \rangle }
\newcommand{\ev}{ {\rm eV} }
\newcommand{\kev}{{\rm keV}}
\newcommand{\mev}{{\rm MeV}}
\newcommand{\gev}{{\rm GeV}}
\newcommand{\tev}{{\rm TeV}}
\newcommand{\mpl}{$M_{Pl}$}
\newcommand{\mw}{$M_{W}$}
\newcommand{\Ft}{F_{T}}
\newcommand{\Zparity}{\mathbb{Z}_2}
\newcommand{\BLambda}{\boldsymbol{\lambda}}
\newcommand{\be}{\begin{eqnarray}}
\newcommand{\ee}{\end{eqnarray}}
	

 %
 %
 %
 %
 %
 %
 %
 %
 %
 %
 %
 %
   \preprint{\vbox{\hbox{ANL-HEP-PR-01-074\\EFI-01-29\\UW-PT-01/20}}}
   \draft                                                                                                                                  
   \title{Radion Mediated Supersymmetry Breaking as a Scherk-Schwarz Theory}
\author{David E Kaplan}
\affiliation{Enrico Fermi Institute,
  The University of Chicago, Chicago, IL 60637, USA;\\
  High Energy Physics Division, Argonne National Laboratory,
  Argonne, IL 60439, USA}
   \author{Neal Weiner}                                                        
  \affiliation{                                                                   
   Department of Physics,                                                      
   University of Washington,                                                   
   Seattle, WA~~98195, USA}                                                    
   \date{\today}                                                               
   \setcounter{footnote}{0}                                                    
   \setcounter{page}{1}                                                        
   \setcounter{section}{0}                                                     
   \setcounter{subsection}{0}                                                  
   \setcounter{subsubsection}{0}                                               
   \begin{abstract}
   	Recently, it has been demonstrated that radion mediated 
supersymmetry breaking gives the same spectrum as 
Scherk-Schwarz supersymmetry breaking, and can be 
interpreted as a dynamical realization of it. We make this connection 
explicit by exhibiting the direct transformation from one theory to 
the other. We then use the extreme UV softness of Scherk-Schwarz 
theories to calculate the one-loop soft masses of matter fields. We 
do not find any cutoff sensitive ``Kaluza-Klein mediated'' 
contributions.
   \end{abstract}
   \maketitle                                                                  


\section{Introduction}

One of the most elegant solutions of the gauge hierarchy problem is 
supersymmetry, in which fields have partners with opposite statistics.
However, superpartners have not been observed.  If supersymmetry is 
realized in nature, it must be broken by some means. 

To this end, many models have been put forward. Recently there has 
been great interest in Scherk-Schwarz supersymmetry breaking 
\cite{Scherk:1979ta}. In this mechanism, bosons and fermions 
belonging to the same supermultiplet are given different global 
transformation properties in a compact space, thus breaking 
bose-fermi degeneracy. 

Recently, Mart\'\i\, and Pomarol \cite{Marti:2001iw} demonstrated that the 
spectra of Scherk-Schwarz theories and those arising from radion 
$F$-components are identical, allowing us to think of the latter 
scenario as a dynamical realization of the former.  For previous
work in this direction, see 
\cite{Ferrara:1989jx,Porrati:1989jk,Ferrara:1994kg,Dudas:1997jn}.

In this letter, we will expand upon the discussion of Mart\'\i\, and 
Pomarol and make the proposed connection explicit by absorbing the 
effects of the radion $F$-term into a field redefinition. Having done so, 
we will show the utility of this connection by calculating UV contributions 
to soft scalar masses in radion mediation \cite{Chacko:2000fn}. In 
particular, by maintaining the manifest UV softness of Scherk-Schwarz 
theories, we do not find the cutoff dependent ``Kaluza-Klein mediated'' 
contributions to soft masses claimed by Kobayashi and Yoshioka 
\cite{Kobayashi:2000ak}.

\section{Radion Mediation and a Field Redefinition}

In this section we show, with a simple field redefinition,
that a theory with a compact extra dimension in which supersymmetry 
is broken by the auxiliary component of the radion superfield is 
equivalent to a theory in which supersymmetry is broken by a
non-trivial winding of some fields ({\it i.e.,} the Scherk-Schwarz 
mechanism).  We first concentrate on bulk gauge fields as they are
relevant to the example in the section 3 and then we discuss
hypermultiplets.  Throughout, we will closely follow the notation
of \cite{Marti:2001iw}.
\vskip -0.2 in

\subsection{Vector Multiplets}
We use four-dimensional $N=1$ superfield notation for extra
dimensional theories.  This elegant tool was developed in
\cite{Arkani-Hamed:2001tb}
with the five-dimensional case generalized to included the radion 
superfield and curved backgrounds in \cite{Marti:2001iw}.  We will be 
working in flat space
with one dimension compactified on a circle with a radius 
described by the radius modulus $R$ and parameterized
by the coordinate $-\pi\leq \varphi < \pi$.  In addition, we impose
an orbifold projection with the identification $\varphi \rightarrow -\varphi $. We use the angular coordinate $\varphi$ to emphasize that 
it has canonical dimension zero. Later, when we set the radion field 
to its vev we will work with the dimension $-1$ coordinate $y=\varphi 
R$.

To use the above described notation we incorporate $R$ into
a chiral superfield:
\begin{equation}
 T=  R + i B_{5} + \theta \Psi^5_R + \theta^{2} F_T\, ,
\end{equation}
where $B_5$ is the fifth component of the graviphoton,
$\Psi^5_R$ is the fifth component of the right-handed
gravitino and $F_T$ is the radion's auxiliary component.

The minimal vector multiplet in five dimensions consists of a 
vector superfield $V$ and a chiral superfield $\chi$:
\begin{align}
    V & = -\theta \sigma^{\mu} \bar{\theta} A_{\mu} -
    i\bar{\theta}^2\mspace{-2mu}\theta\lambda_{1} +
    i\theta^2\mspace{-2mu}\bar{\theta}\bar{\lambda}_{1} +
    \frac{1}{2} \bar{\theta}^2\mspace{-2mu}\theta^2 D\, ,\nonumber \\
    \chi &= \frac{1}{\sqrt{2}} \left(\Sigma + i A_{5}\right) +
   \sqrt{2} \theta \lambda_{2} + \theta^{2} F_\chi\, .
\end{align}
We can now write down the action of a five-dimensional Abelian 
gauge multiplet coupled to the radion \cite{Marti:2001iw}:
\be
\label{GaugeAb}
    S_5 = \int d^{4}\! x d\! \varphi\, 
{\biggl[} \frac{1}{4g^2_5} \int d^2\mspace{-2mu}\theta\, T
        W^{\alpha}W_{\alpha} + \text{h.c.} \\ \nonumber + \frac{2}{g^2_5} \int
         d^4\mspace{-2mu}\theta\: \frac{1}{(T +
          T^{\dag})} \left( \partial_\varphi V - \frac{1}{\sqrt{2}} (\chi +
            \chi^{\dag})\right)^2 {\biggr]}\, .
\ee
Note, this action is invariant under the full five-dimensional gauge
transformation $V\rightarrow V+\Lambda+\Lambda^{\dag}\, ,
\chi\rightarrow \chi+\sqrt{2}\partial_5\Lambda$ and can be shown to 
give the correct component-field action.

Now let us assume the radion has a non-zero auxiliary component such
that $\langle T \rangle = R + \theta^2 F_T$.  After eliminating the
other auxiliary fields via their equations of motion, replacing the
radion with it's vacuum expectation value and rescaling the fields
$\Sigma \rightarrow R \Sigma$, $ \lambda_2 \rightarrow - i R \lambda_2$,
and replacing the coordinate $\varphi$ with $ y/R$, we have:
\be
    \label{component}
S_5 = \frac{1}{g^2_5}\int d^{4}\! x d\! y\, \sqrt{-g} \biggl[ 
-\frac{1}{2}\partial_{M}\Sigma
    \partial^{M} \Sigma - \frac{1}{4} F_{MN} F^{MN} 
\\ \nonumber - i \lambda_i \sigma^{\mu} \partial_{\mu} \bar{\lambda}_{i}
+ \frac{1}{2} \lambda_i \epsilon_{ij} \partial_y \lambda_j + {\rm 
h.c.} \\ \nonumber
- \left( \frac{F_T}{4 R}{\lambda}_{1} {\lambda}_{1} 
+ \frac{F_T^{\dagger}}{4 R}{\lambda}_{2} {\lambda}_{2} \right)
+ {\rm h.c.}
    \biggr]\, ,
\ee
where $\sigma^m = (1, \vec \sigma)$, $\overline{ \sigma}^m = (1, -\vec \sigma)$.
We use two-component spinor for simplicity of notation when we couple the
theory to boundary fields below

The action for the non-Abelian theory appears in \cite{Marti:2001iw} and will 
not be presented here.  The arguments below are presented for the case
of an Abelian theory and apply equally well to the non-Abelian case.


Now we show that a theory with supersymmetry
breaking by a radion is equivalent to a theory where supersymmetry
is broken explicitly by boundary conditions in the fifth dimension
({\it i.e.,} the Scherk-Schwarz mechanism).  This can be done simply by
a field redefinition equivalent to an $y$-dependent SU(2)$_R$ transformation.

We perform the following field redefinition on the action in 
(\ref{component}) \cite{symplectic}:
\begin{equation}
\label{trans}
   \begin{pmatrix} \lambda_1 \cr \lambda_2\cr\end{pmatrix}\ , 
   \rightarrow
   e^{-i F_T y \sigma^2/2 R}
   \begin{pmatrix} \lambda_1 \cr \lambda_2\cr\end{pmatrix} \ ,
\end{equation}
where we take $F_T$ to be real.
This transformation is similar to the SU(2)$_R$ component of the 
field redefinition in \cite{Pomarol:1998sd} (for generalizations, see \cite{Bagger:2001qi}). 
Note that the only term in (\ref{component}) which is not invariant under this 
transformation is the fifth component of the gaugino kinetic term.  
This term transforms as
\begin{equation}
\frac{1}{2 g^2_5} \lambda_{i} \epsilon_{ij} \partial_{y} \lambda_{j}
 \rightarrow 
\frac{1}{2 g^2_5} \lambda_{i} \epsilon_{ij} \partial_{y} \lambda_{j}
+ \frac{F_T}{4 g^2_5 R} \lambda_{i} \lambda_{i}.
\end{equation}
Here it is chosen to completely remove the Majorana mass terms for the 
gaugino fields at the expense of giving them a non-trivial winding around
the compact dimension.  The winding is precisely the Scherk-Schwarz 
mechanism for breaking supersymmetry.

The above transformation can be generalized to complex $F_T$ by replacing the
phase as 
\begin{equation}
F_T \sigma^2 y/2R\rightarrow {\bf F}_T \cdot {\bf\sigma} y/2R
\end{equation}
where ${\bf F}_T=\{ -{\rm Im}F_T, {\rm Re}F_T,0\}$.

The winding of the gauginos effects the gaugino coupling to charged matter 
living at the boundary $y=0$ and their identified points $y=2\pi R n$.  
The coupling of the bulk gauge multiplets to charged fields $Q$ living at the boundary
$y=0$ appears as
\begin{equation}
    S_{\rm boundary} = \int d^{5}\! x\, 
        \left[ \int d^4\mspace{-2mu}\theta\, 
        Q^{\dagger}e^V Q \right]\delta(y)\,.
\end{equation}
These couplings, and the field content at the boundary, need only preserve
an $N=1$ component of the supersymmetry.  Now instead of a compact direction, 
let us treat the fifth dimension as infinite and impose a periodicity condition 
on the Lagrangian, in which case the above delta function would become a sum of 
delta functions $\sum_n \delta(y-2\pi R n)$.  The the fields at $y=0$ couple to the 
gaugino $\lambda_1$  as one would expect in a normal four-dimensional $N=1$ theory.
At $y=2\pi R n$ however, the boundary fields couple to the linear combination 
$\cos{\omega n} \lambda_1 + \sin{\omega n} \lambda_2$ where
\begin{equation}
\omega = \pi | F_T |.
\end{equation}
Because supersymmetry breaking is due to non-trivial winding of fields, loop 
corrections to soft parameters will be physically cut off by the compactification
scale and thus rendered finite.  This will be important in Section 3 when we
do explicit calculations in one picture to gain information about the other.

\subsection{Hypermultiplets}
We now turn our attention to hypermultiplets. We again use the 
formalism developed by \cite{Marti:2001iw}. We will not consider 
fields transforming under gauge symmetries, but to include them does 
not change the features of the arguments here.

The coupled radion-hypermultiplet action is given by \cite{Marti:2001iw}
\be
S=\int d^{4}x\! d\varphi \biggl( \int d^{4}\theta 
\frac{(T+T^{\dagger})}{2}\left( \Phi^{\dagger} \Phi + \Phi^{c} \Phi^{c 
\dagger} \right)\nonumber \\
+ \int d^{2}\theta \Phi^{c}\partial_{\varphi}\Phi \biggr).
\ee
If we include a radion $F$-term, the $F$ terms of the hypermultiplet 
fields are
\be
F_{\phi}&=&\frac{1}{R}(\partial_{\varphi} \phi^{c*}-\frac{F_{T}}{2}\phi) \\
F_{\phi^{c *}}&=&-\frac{1}{R}(\partial_{\varphi} 
\phi-\frac{F_{T}}{2}\phi^{c*}).
\ee
Defining $\overline \phi = (\phi \> \phi^{c\dagger})$, rescaling 
$\overline \phi \rightarrow \overline \phi / \sqrt{R}$ and
$\varphi\rightarrow y/R$ and taking $F_T$ to be real i
gives a scalar potential of 
\be
V(\overline \phi) = \overline \phi^{\dagger} 
\begin{pmatrix}
 -\partial_{y}^{2}+F_{T}^{2}/4 R^2 & - F_{T}/R \partial_{y} \cr
F_{T}/R \partial_{y} & -\partial_{y}^{2} + F_{T}^{2}/4 R
\end{pmatrix}
\overline \phi \\ \nonumber 
= \overline \phi^{\dagger} \left[(-\partial_{\varphi}^{2}+F_{T}^{2}/4) {\bf 
1} + i F_{T} \partial_{\varphi} \sigma_{2} \right] \overline \phi
\ee
We now redefine $\overline \phi$ by rotating it 
\be
\overline \phi \rightarrow 
   e^{-i F_T y \sigma^2/2 R}
\overline \phi.
\ee
The potential is now
\be
V(\overline \phi') = 
\overline \phi^{\dagger} \begin{pmatrix}
	-\partial_{\varphi}^{2} &
	0 \cr 0 & 
	-\partial_{\varphi}^{2}
\end{pmatrix} \overline \phi.
\ee
With the same rotation as was required by the gauge sector, the mass
terms vanish.  Note the same generalization to complex $F_T$ works
here as well.

\section{Radion Mediation with Bulk Gauge Fields}
If gauge fields propagate in the bulk, a radion $F$-term can
generate tree-level gaugino masses in realistic theories, {\it i.e.,}
those in which the radius is stabilized \cite{Chacko:2000fn}. 
It is thus a natural framework in which to realize the 
gaugino mediated supersymmetry breaking scenario 
\cite{Kaplan:1999ac,Chacko:1999mi}.  However, in \cite{Kobayashi:2000ak}, 
it was argued that a radion $F$-term gave rise to cutoff dependent 
contribution to scalar masses. 

Recently, there has been much discussion about the softness of 
Scherk-Schwarz supersymmetry breaking 
\cite{Antoniadis:1998zg,Antoniadis:1998sd,Barbieri:2000vh,Arkani-Hamed:2001mi,Ghilencea:2001ug,Delgado:2001ex,Contino:2001gz,Nomura:2001ec,Weiner:2001ui,Kim:2001gk,Barbieri:2001dm,Masiero:2001im,Delgado:2001xr}. 
Hard cutoffs generally do not preserve the important features of the theory 
(namely locality), and thus can give rise to artificial cutoff dependence
in physical quantities.

Because we
now can utilize the equivalence between a radion 
$F$-component vev and a Scherk-Schwarz compactification, we can 
calculate one-loop contributions to soft masses without any cutoff. 
Since we wish to preserve locality in our calculations, it will
be easiest if we work in mixed position/momentum space 
\cite{Arkani-Hamed:1999za}.

Let us review this formalism.  We will be working in a five 
dimensional theory. Since issues of locality are essential to 
understanding the physics of the setup, we leave the fifth 
coordinate $y$ explicit, while we Fourier transform the Minkowski 
dimensions to momentum space.

In infinite space, the propagator for a fermion is 
\be
\nonumber
G(k_{4},y) &=& \int d k_{5} \frac{\not k_{4}+\gamma_{5} 
k_{5}}{k_{4}^{2}+k_{5}^{2}} e^{-i k_{5} y}\\
\label{eq:infprop}
&=& \frac{\not k_{4}}{2 k_{4}}e^{-k_{4} |y|}+ i \gamma_{5} 
\partial_{y} \left(\frac{e^{-k_{4}|y|}}{2 k_{4}}\right)\\
\nonumber
&=&\frac{\not k_{4}}{2 k_{4}}e^{-k_{4} |y|}- i \gamma_{5} 
(\theta(y)-\theta(-y))\frac{e^{-k_{4}|y|}}{2}.
\ee
Note that here we have analytically continued the propagator to be a 
function of Euclidean momentum $k_{4}$. For calculation of the soft 
masses, the second term, proportional to $\gamma_{5}$, will be 
irrelevant, and we will henceforth neglect it.

Of course, this is the propagator in infinite space. What we are 
really after is the propagator in a compact space, and, moreover, the 
propagator in a space with a Scherk-Schwarz compactification. This is 
easily achieved using (\ref{eq:infprop}).

If we have a source $J$ at $y=0$, then the propagator amplitude to 
propagate from $y=0$ to $y=0$ is just $G(k,0)$. However, we can also 
propagate from $y=0$ to $y= 2 n \pi R$, with an amplitude $G(k,2 n \pi 
R)$. We must be careful, however, because the periodicity conditions 
are non-trivial, and thus the source on the brane at $y=2 n \pi R$ 
couples to a different linear combination of $\lambda_{1}$ and 
$\lambda_{2}$, given by the $SU(2)_{R}$ transformation. Thus the true 
amplitude to propagate from a brane to itself, summing all windings is
\begin{align}
\label{eq:gprop}
\overline G(k,0) &= \sum_{n} G(k, 2 n \pi R) \cos (\omega n)\\ 
\nonumber
&= 
\frac{\not k \sinh(2 \pi k R)}{4 k(\sinh^{2}(k \pi R) 
\cos^{2}(\frac{\omega}{2}) + \cosh^{2}(k \pi R) \sin^{2}(\frac{\omega}{2}))} .
\end{align}
Notice that if we analytically continue $k$ back to Minkowski 
momentum, the propagator has poles for
\be
k = \frac{\omega}{2 \pi R} \pm \frac{n}{R}. 
\ee
Thus the poles in the propagator are precisely those of the spectrum 
in section 2. By summing over winding modes for an 
infinite space propagator, we have summed over the discrete spectrum 
of the compact space Kaluza-Klein tower.

\subsection{One loop contributions to soft masses}

With our approach explicit, we can calculate the contributions to soft 
masses. The gaugino loop contributions are given by:
\begin{eqnarray}
\label{eq:gauginocon}
(-1) \times (\sqrt{2} g_{5})^{2} C_{2}(G)\int \frac{d^{4}k}{(2 \pi)^{4}}
Tr[P_L \frac{1}{\not k} \overline G(k,0)].
\ee
There are other diagrams contributing to the soft 
masses at one loop. The gaugino does not appear in any of these diagrams, however, 
and thus these diagrams are not aware of the supersymmetry 
breaking parameter $\omega$. Nonetheless, 
these must cancel the contribution in 
(\ref{eq:gauginocon}) in the limit of no supersymmetry breaking. No 
supersymmetry breaking corresponds to $F_{T}=0$, or, equivalently, 
$\omega = 0$. Thus, we can reevaluate (the negative of)
(\ref{eq:gauginocon}) with $\omega$ set to 0, and this gives the 
contribution of the remaining diagrams. Integrating 
the total contribution to the 
soft masses is up to a scale $\Lambda$ is
\begin{align}
\label{eq:softmass}
m_{\phi^{2}}(\Lambda)&=\\
\nonumber 
& \int_{0}^{\Lambda} d^{4}k 
	\frac{g_{5}^{2} C_{2}(G)k^{2} \coth(k \pi R)}{4 \pi^{2} (\sinh^{2}(k \pi R) \cot^{2}(\frac{\omega}{2}) +\cosh^{2}(k \pi R))}.
\end{align}
Because of the exponential suppression in the $k\gg R^{-1}$ 
limit, this is finite in the UV, giving a total contribution
\be
\label{eq:softmass2}
m_{\phi^{2}}
&=& \frac{g_{5}^{2}  C_{2}(G) }{16 \pi^{5} R^{3}}
	\left(2 \zeta(3)-{\rm Li}_{3}
	(e^{i \omega})-{\rm Li}_{3}(e^{-i\omega})\right).
\end{eqnarray}
Note that this is precisely the same formula found in \cite{Delgado:1998qr} 
in the context of \tev-sized extra dimensions.

Since we are assuming the dimension is small and supersymmetry 
breaking is weak, we expand about small $\omega$ making the identification
$M_{1/2} = \omega /2 \pi R \equiv \omega / M_c$, and find
\begin{eqnarray}
	\label{eq:totalmass}
m_{\phi^{2}}&\approx&\frac{g_{4}^{2} C_{2}(G) M_{1/2}^{2}}{4 \pi^{2}}(3
	+2 \log(M_c / M_{1/2})).
\end{eqnarray}
While this is a complete one-loop calculation, we have not included 
the fact that both $g_{5}$ and $M_{1/2}$ run as functions of energy 
as well (note, $M_{1/2}$ does not run above the compactification
scale).  It would have been more accurate, in fact, to 
replace $g_{5}^{2}$ in (\ref{eq:softmass}) with $g_{5}^{2}(k)$ and 
$M_{1/2}(k)$. We can study the running of $m_{\phi}^{2}$ by 
differentiating with respect to the cutoff. For 
 $\omega R^{-1} \ll \Lambda \ll R^{-1}$, we have
\be
\Lambda \frac{d m_{\phi^{2}}}{d \Lambda}= \frac{g_{5}^{2} C_{2}(G) \omega^{2}}{16 
\pi^{5} R^{3}} = \frac{g_{4}^{2} C_{2}(G) M_{1/2}^{2}}{2 \pi^{2}},
\ee
we we immediately recognize this as the renormalization group 
equation for the running of the soft mass squared.

Given this, we can identify the $\log(R^{-1}/M_{W})$ piece of 
(\ref{eq:totalmass}) as the ordinary gaugino 
mediated contribution. All other pieces are both finite and small.

One might be concerned that if $g_{5}$ began power law running it 
could change this and introduce cutoff dependence. There are two 
simple reasons why this is not the case: first, quantum corrections 
actually drive $g_{5}$ {\em down}, thus this is an upper limit on the 
UV contributions. Secondly, even if there were a positive power law 
piece, the integrand in (\ref{eq:softmass}) goes as $e^{-k 2 \pi R}$ 
for $k > R$, and thus even with power law growth, these contributions 
damp off exponentially.

There is a simple physical interpretation for this: divergences are 
associated with contracting loops to a point. Here, supersymmetry 
violating contributions must sample the entire space, that is, they 
must wind at least once around the extra dimensions. These 
contributions cannot be contracted to a point and hence will not be 
divergent.

\section{Conclusion}

We have shown that a five-dimensional theory with a radion $F$ term is 
equivalent to a theory with Scherk-Schwarz boundary conditions up to a
field redefinition.  This makes more concrete the statement by
Mart\'\i\, and Pomarol that radion mediated supersymmetry breaking is
simply a dynamical realization of Scherk-Schwarz breaking.  The
equivalence makes the calculation of scalar masses easier and we find 
contributions from Kaluza-Klein modes to be unimportant.

It is interesting to note that a radion $F$ term translates into a 
specific Scherk-Schwarz theory.  For example, the boundary conditions
in \cite{Pomarol:1998sd} require an additional twist along the direction
of a different symmetry SU(2)$_H$.  It would be interesting to see if 
dynamical versions of such theories could be realized.

\vskip 0.25in
{\bf Acknowledgements} 
\vskip 0.15in
The authors thank Gia Dvali, Ann Nelson and Alex Pomarol for reading a 
draft of the paper and for useful discussions and the 
Aspen Center for Physics where this work was completed. 
This work was partially supported by the DOE
under contracts DE-FGO3-96-ER40956, DE-FG02-90ER-40560 and W-31-109-ENG-38. 
\bibliography{radio}
\bibliographystyle{apsrev}
\end{document}